\documentclass{osa-article}

\journal{osajournal}

\articletype{Research Article}

\begin{document}

\title{Tunable Photon blockade with single atom in a cavity under electromagnetically induced transparency}

\author{Jing Tang,\authormark{1} Yuangang Deng,\authormark{1,*} and Chaohong Lee\authormark{1,2,3,*}}

\address{\authormark{1}Guangdong Provincial Key Laboratory of Quantum Metrology and Sensing $\&$ School of Physics and Astronomy, Sun Yat-Sen University (Zhuhai Campus), Zhuhai 519082, People's Republic of China\\
\authormark{2}State Key Laboratory of Optoelectronic Materials and Technologies, Sun Yat-Sen University (Guangzhou Campus), Guangzhou 510275, People's Republic of China\\
\authormark{3}Synergetic Innovation Center for Quantum Effects and Applications, Hunan Normal University, Changsha 410081, People's Republic of China}

\email{\authormark{*}dengyg3@mail.sysu.edu.cn} 
\email{\authormark{*}lichaoh2@mail.sysu.edu.cn}



\begin{abstract}
We present an experimental proposal to achieve a strong photon blockade by employing electromagnetically induced transparency (EIT) with single alkaline-earth-metal atom trapped in an optical cavity. In the presence of optical Stark shift,  both second-order correlation function and cavity transmission exhibit asymmetric structures between the red and blue sidebands of the cavity. For a weak control field, the photon quantum statistics for the coherent transparency window (i.e. atomic quasi-dark state resonance) are insensitive to the Stark shift, which should also be immune to the spontaneous emission of the excited state by taking advantage of the intrinsic dark-state polariton of EIT. Interestingly, by exploiting the interplay between Stark shift and control field, the strong photon blockade at atomic quasi-dark state resonance has an optimal second-order correlation function $g^{(2)}(0)\sim10^{-4}$ and a high cavity transmission simultaneously. The underlying physical mechanism is ascribed to the Stark shift enhanced spectrum anharmonicity and the EIT hosted strong nonlinearity with loss-insensitive atomic quasi-dark state resonance, which is essentially different from the conventional proposal with emerging Kerr nonlinearity in cavity-EIT. Our results reveal a new strategy to realize high-quality single photon sources, which could open up a new avenue for engineering nonclassical quantum states in cavity quantum electrodynamics.
\end{abstract}

\section{Introduction}
Single photon state is a fundamental concept in quantum physics that has been actively explored for their potential applications in quantum computing~\cite{Bennett00, Knill01,  Duan04, Pieter07, Buluta11}, quantum information processing~\cite{Duan01, Scarani09, Brien09, Giovannetti2011}, and strongly correlated man-body quantum phenomena~\cite{Chang08,Chang14,Hartmann_2016}. One of the critical underlying techniques for generating single photon emission is to achieve photon blockade (PB), which implies that the excitation of the first photon will block the transmission of the later arriving photons, resulting in an orderly output of photon stream with sub-Poissonian statistics~\cite{schmidt1996giant,Imamoglu97}. Up to now, the typical mechanism for realizing PB includes conventional PB~\cite{Birnbaum2005} and unconventional PB~\cite{Liew10,Tang15}, which rely on the strong  coupling induced strong energy-spectrum anharmonicity in the emitter-cavity system and the quantum interference between the different excitation pathways, respectively. The above two mechanisms for generation of strong PB have been investigated theoretically and experimentally on various platforms based on the Jaynes-Cummings model of a two-level system, e.g., quantum dot- and atom-cavity systems~\cite{Hennessy07, Kai15, Fink08, Faraon08, Majumdar2012, Tang15}, optomechanical systems~\cite{Rabl11,Nunnenkamp11,Wang15,Sarma18}, and waveguide- and circuit-QED systems~\cite{Mirza16,  Mirza17, Gheeraert18, Zheng11, Mirza2016, Hoffman11,Lang11, Liu14}. In particular, unconventional PB has been experimentally demonstrated in single quantum dot-cavity system and two coupled superconducting resonators~\cite{Snijders18,Vaneph18} by utilizing the constructive quantum interference to suppress the two-photon excitation. However, our recent work shows that the quantum interference mechanism for generating strong PB    [$g^{(2)}(0)\sim0$] is still a challenge for the weak atom-cavity coupling regime~\cite{Tang19}, which indicates that sufficient energy-spectrum anharmonicity is necessary for most of the systems. In general, multiphoton excitation plays an important role in the system with weak energy-spectrum anharmonicity, although the two-photon excitation could be completely suppressed by quantum interference for unconventional PB.

On the other hand, electromagnetically induced transparency (EIT) with its intriguing loss free dark-state polariton (DSP)~\cite{PhysRevLett.84.5094} has been observed in atom-cavity systems~\cite{Harris1990, Lukin01, Kang03, Fleischhauer05, Eisaman05}, which offers even further opportunities for realizing a large and tunable nonlinearity from the control of light with light, i.e., giant Kerr effect~\cite{schmidt1996giant,Imamoglu97}. Moreover, remarkable progress on realization of single atom EIT in a high-finesse optical cavity~\cite{Mucke10,Souza13,Tanji-Suzuki1266} could  allow for novel applications ranging from quantum computing~\cite{Albert2011,Chen768,Kampschulte10, Peyronel12, Holleczek16, Li2018} to quantum simulation~\cite{Barrett13} and coherent quantum networks~\cite{Kimble08, Ritter12}. Thus, it is natural to wonder whether strong PB can be realized in single atom cavity-EIT by employing the DSP, which is insensitive to the loss of the atomic excited state and possesses a significantly enhanced coherent lifetime of quantum state~\cite{Wu08}. An affirmative answer in exploring this question will significantly enhance our exciting insights in quantum physics, as well as potential novel applications in quantum information science, where an ideal single photon source is achieved by combining the advantages of strong nonlinearity and DSP in cavity-EIT.

In this work, we investigate the realization of strong PB by utilizing an optical Stark shift in single alkaline-earth-metal atom EIT placed inside a high-finesse cavity. It is shown that the combination of EIT hosted strong nonlinearity and Stark shift enhanced energy-spectrum anharmonicity gives rise to the strong PB, which, in the mean time, maintains a high transmission ascribed to the atomic quasi-dark state resonance in EIT. Compared with the conventional proposal of generating strong interacting photons in cavity coupled condensates~\cite{Imamoglu97}, the PB in our single atom-cavity-EIT system depends on the optimal parameters of the Stark shift and Rabi frequency of the classical control field. Our results reveal that a large enough Stark shift and control field are not always beneficial for strong PB, which   is different from  the well-known mechanism for generating PB with giant Kerr nonlinearity in cavity-EIT. In particular, the optimal second-order correlation function [$g^{(2)}(0)\sim 10^{-4}$] in our model can be reduced more than $45$ times in magnitude compared with the extensively studied single four-level atom-cavity-EIT with the effective Kerr nonlinearity at a moderate Stark shift~\cite{PhysRevA.61.011801,Rebic_1999,rebic2002photon,bajcsy2013photon}.

\section{Model and experimental feasibility}

\begin{figure}[ht]
\centering
\includegraphics[width=0.8\columnwidth]{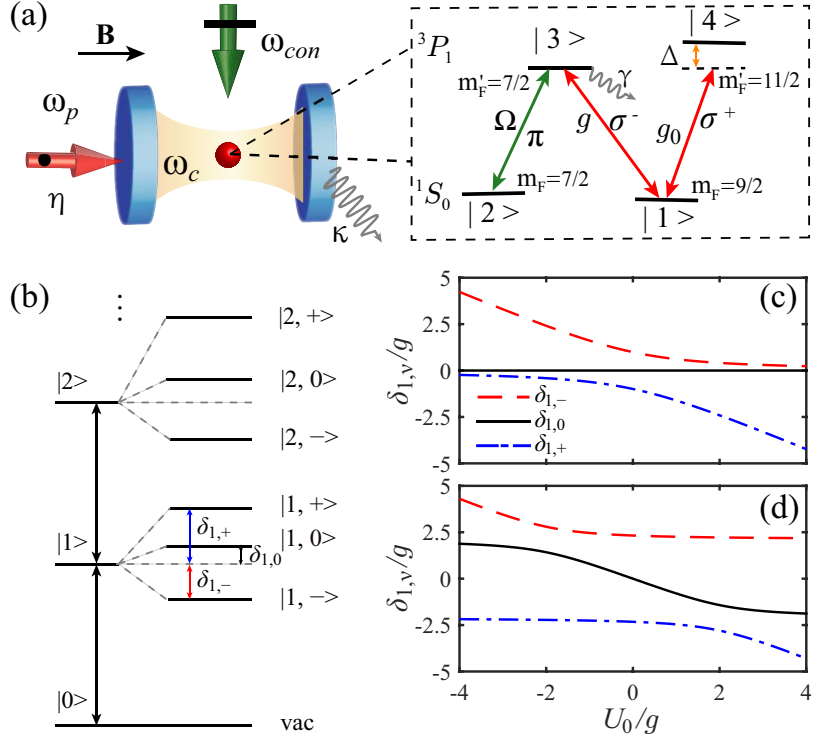}
\caption{(color online). (a) Schematic diagram of the relevant transitions in cavity-EIT system. A single $^{87}$Sr four-level atom is trapped inside a high-finesse optical cavity. The bias field ${\bf B}$ parallels to the cavity axis, which defines the quantization axis and generates a Zeeman splitting $\hbar \Delta$ between the magnetic sublevels $|3\rangle$ and $|4\rangle$ of $^3P_1$. The cavity is driven by a weak $\sigma$-polarized laser field  (given by the superposition of $\sigma^+$ and $\sigma^-$ polarization)  and the atom is pumped by a $\pi$-polarized classical control field orthogonal to the cavity axis. (b) The typical energy spectrum for optical Stark shift mediated cavity-EIT. The dressed-state splitting $\delta_{1,-} $ (the red dashed line), $\delta_{1,0} $ (the black line) and $\delta_{1,+} $ (the blue dotted line) as a function of the optical Stark shift $U_0$ for $\Omega/g=0.01$ (c) and $\Omega/g=2.1$ (d), respectively. } \label{scheme}
\end{figure}%

We consider a single alkaline-earth-metal (like) atom inside a high-finesse cavity subjected to a bias magnetic field ${\bf B}$ along the cavity axis (quantization axis). Figure.~\ref{scheme}(a) shows the level structure of the electronic
$^1S_0$ ($F=9/2$) state to $^3P_1$ ($F'=11/2$) state transition for the $^{87}$Sr atom, in which the four relevant magnetic Zeeman sublevels with two ground states $|F=9/2, m_{F}=9/2\rangle$ ($|1\rangle$) and $|F=9/2, m_{F}=7/2\rangle$ ($|2\rangle$), two long-lived electronic orbital states $|F'=11/2, m'_{F}=7/2\rangle$ ($|3\rangle$) and $|F'=11/2, m'_{F}=11/2\rangle$ ($|4\rangle$) are included. 
In our configuration, the cavity field only supports a $\sigma$-polarized (orthogonal to the bias magnetic field ${\bf B}$) photon, which can be decomposed into $\sigma^-$- and $\sigma^+$-polarized modes~\cite{specht2011single}. 
The dipole-forbidden singlet $^1S_0$ to triplet $^3P_1$ optical transition is coupled to a $\sigma$-polarized cavity field with a natural narrow linewidth of $7.5$ kHz~\cite{Norciae1601231}. 
During this coherent process, the magnetic quantum numbers of the electronic states induced by the $\sigma^+$- and $\sigma^-$-polarized cavity field satisfy $m'_{F}-m_{F}= \pm 1$. 
Specifically, the single atom is initially prepared in the state $|1\rangle$. 
The transition $|1\rangle \leftrightarrow |3\rangle$ at a wavelength $698$ nm is resonantly driven by the $\sigma^-$-polarized optical cavity, corresponding to the single atom-cavity coupling $g$ and the intrinsic cavity decay rate $\kappa$. 
Besides the resonant transition $|1\rangle \leftrightarrow |3\rangle$, the $\sigma^+$-polarized cavity field also drives the far-resonance atomic transition $|1\rangle \leftrightarrow |4\rangle$ with the single atom-cavity coupling $g_0$ and the atom-cavity detuning $\Delta$. 
Here the tunable $\hbar\Delta$ denotes the Zeeman splitting between $|F'=11/2, m'_{F}=7/2\rangle$  and $|F'=11/2, m'_{F}=11/2\rangle$ of the $^3P_1$ state.
A tunable classical $\pi$-polarized (along the ${\bf B}$ field) control field propagating orthogonal to the cavity axis drives the resonantly atomic transition $|2\rangle \leftrightarrow |3\rangle$ with Rabi frequency $\Omega$. In addition, the cavity is driven by a weak $\sigma$-polarized (orthogonal to the ${\bf B}$ field) laser field with frequency $\omega_p$ and  the driven amplitude $\eta$ that related to the input laser power $P$ and optical-cavity decay rate $\kappa$ by $\eta=\sqrt{2P\kappa/\hbar
\omega_p}$, which yields the cavity-light detuning $\Delta_c=\omega_c-\omega_p$ with $\omega_c$, being the bare cavity frequency.

In the limit of large atom-cavity detuning of $|g_0/\Delta|\ll1$, the atomic excited state $|4\rangle$ can be adiabatically eliminated, which yields a $\Lambda$-level scheme for generating cavity-EIT with the tunable optical Stark shift $U_0=-g_0^2/\Delta$. By using the rotating wave approximation, the relevant Hamiltonian of the single atom-cavity-EIT system takes the form as
\begin{align}\label{Hamiltonian}
{\hat{\cal H}}/\hbar =& g (\hat{a}^\dag \hat{\sigma}_{13}+\hat{a} \hat{\sigma}_{31})+\Omega(\hat{\sigma}_{23}+\hat{\sigma}_{32}) +  U_0 \hat{a}^\dag\hat{a} \hat{\sigma}_{11}  \nonumber \\
&+\Delta_c \hat{a}^\dag\hat{a} +\Delta_c\hat{\sigma}_{33}+\Delta_c\hat{\sigma}_{22} +\eta(\hat{a}^\dag  + \hat{a}),
\end{align}%
where $\hat{a}^\dag$ ($\hat{a}$) is the creation (annihilation) operator of the single mode cavity, $\hat{\sigma}_{ij}$ are the atomic spin projection operators with $i, j=1, 2, 3$ labeling the internal states of the atom. For simplicity, the two-photon detuning of the Raman process is fixed equal to the cavity-light detuning $\Delta_c$ by tuning the control field.

Compared with the experimental studies on single atom-cavity-EIT in Ref.~\cite{Mucke10}, the additional optical Stark shift in Eq.~(\ref{Hamiltonian}) plays an important role in changing the energy spectrum of the cavity-EIT system, especially controlling the quantum statistical properties of the transmission photons. In addition, the term of $U_0 \hat{a}^\dag\hat{a} \hat{\sigma}_{11}$ in our model could facilitate more for the emergence of strong PB at a moderate strength of $U_0$ in contrast to the proposals in single four-level atom-cavity-EIT with the Stark shift term of $U_0 \hat{a}^\dag\hat{a} \hat{\sigma}_{22}$~\cite{PhysRevA.61.011801,Rebic_1999,rebic2002photon,bajcsy2013photon}, as we shall see below.

To generate the large $U_0$, the condition of $|g_0/g|^2\gg 1$ should be satisfied. Therefore, the matrix element of atomic transition $|1\rangle \leftrightarrow |3\rangle$ should be much smaller than $|1\rangle \leftrightarrow |4\rangle$. In our configuration in Fig.~\ref{scheme}(a), we find that the atom-cavity coupling strength ratio  between  the $\sigma^+$ and $\sigma^-$ polarization of the cavity field is $|g_0/g|^2=165/3\gg 1$ by calculating the relevant transition matrix elements. In fact, the analogous proposal by utilizing dipole-forbidden transition has been realized experimentally in cavity enhanced magnetically induced optical transparency on $^{88}$Sr atomic ensembles in Ref.~\cite{PhysRevLett.118.263601}. Moreover, we should note that the elaborately selective long-lived state in our model can also be employed by utilizing the clock transition, owing to the unique energy-level structures for alkaline-earth-metal atoms~\cite{PhysRevLett.117.220401,kolkowitz2017spin,bromley2018dynamics}.

\section{Energy spectrum}

To investigate the underlying physical mechanism for photon quantum statistics with combining the cavity-EIT induced nonlinearity and Stark shift, we calculate the energy spectrum of the system by diagonalizing the Hamiltonian in Eq.~(\ref{Hamiltonian}) with ignoring the weak driving field of the cavity. The total excitation numbers for the single $\Lambda$-level atom trapped in the cavity-EIT system is conserved with $\eta=0$. Then, the accessible Hilbert space can be restricted to Fock states with the type $|n,N_1, N_2, N_3\rangle$, where $N_i(i = 1, 2, 3) $ represents the number of atoms trapped in $|i\rangle$ state and $n$ denotes the photon number for the cavity. Explicitly, the relevant three spanned states are $|n,1, 0, 0, \rangle$, $|n-1,0, 1 , 0\rangle$, and $|n-1,0, 0 , 1\rangle$. Therefore, the corresponding block can be written in matrix form in the basis ${\hat{\cal H}}\Psi={\cal M}\Psi$ with $\Psi=[|1, 0, 0, n\rangle, |0, 1 , 0, n-1\rangle, |0, 0 , 1, n-1\rangle]^T$. The matrix ${\mathcal {M}}$ is obtain as
\begin{align}\label{matrix}
{\mathcal {M}}=\left(\begin{array}{ccc}
n\Delta_c+nU_0 & 0 & g\sqrt{n}\\
0 & n\Delta_c &  \Omega \\
g\sqrt{n} &\Omega & n\Delta_c \\
\end{array}\right)
\end{align}%
by diagonalizing the matrix in Eq.~(\ref{matrix}), the energy spectrum for the nonzero photon excitation recasts into three branches. Fig.~\ref{scheme}(b) shows the typical anharmonicity ladder of the energy spectrum. The asymmetric energy splitting of the $n$th dressed states induced by the Stark shift for the higher- ( $|n,+\rangle$ ), middle- ( $|n,0\rangle$ ) and lower- ( $|n,-\rangle$ ) branches are labeled as $\delta_{n,+}$, $\delta_{n,0}$, and $\delta_{n,-}$, respectively. For $U_0=0$, the middle-branch splitting $\delta_{n,0}=0$, which exhibits a DSP that denotes atomic dark-state resonance. It is obvious that the DSP vector can be written as
\begin{align}\label{DSP}
|n,0\rangle = \beta_+ |n\rangle|1\rangle + \beta_0 |n-1\rangle|3\rangle + \beta_- |n-1\rangle|2\rangle,
\end{align}
with $\beta_+=-\Omega/\sqrt{\Omega^2+g^2n}$, $\beta_-=g\sqrt{n}/\sqrt{\Omega^2+g^2n}$, and $\beta_0=0$. However, this dark-state resonance should be strictly called ``quasi-dark-state resonance'' in the presence of nonzero $U_0$, where the DSP is broken with $\beta_0\neq 0$.

Figures.~\ref{scheme}(c) and \ref{scheme}(d) show the dressed-state splitting $\delta_{1,\nu}$ as a function of Stark shift $U_0$ for weak ($\Omega/g=0.01$) and strong ($\Omega/g=2.1$) control fields by fixing $g/\kappa = 4$, respectively, where $\nu=\pm1,0$ represents the upper, lower, and middle helicity branches, respectively. An asymmetry structure of the dressed-state splitting between the red and blue light-cavity detuning is observed, where the result is analogous to our recent study on the Stark shift mediated Jaynes-Cummings model~\cite{Tang19}. Moreover, the $U_0$-dependent lower branch splitting $\delta_{1,-}$ is significantly enhanced with the increasing negative $|U_0|$, while gradually saturated to a constant for positive $U_0$, vice verse for the case of the higher branch splitting $\delta_{1,+}$.

Interestingly, the middle-branch splitting at atomic quasi-dark-state resonance is insensitive to $U_0$ with roughly zero shift ($\delta_{1,0}\sim 0$) in the presence of a weak control field. But, for a large control field, the value of $\delta_{1,0}$ enhances rapidly with the increasing of both positive and negative Stark shift $|U_0|$. As expected, the observed higher anharmonic ladder of the energy spectrum could facilitate the generation of strong nonclassical photons by employing the enhanced nonlinearity in Stark shift mediated cavity-EIT. In the following, we will show the strong PB induced by combining optical Stark shift $U_0$ and the loss-insensitive atomic quasi-dark-state resonance in the EIT-cavity system.

\section{Numerical results}

The quantum statistical properties of photons in the cavity are characterized by numerically solving the quantum master equation using Quantum Optics Toolbox~\cite{Tan99}, which takes into account the dissipation of photons and atom. 
The atom-cavity system can be described by the density matrix $\rho$ for the cavity mode and the atomic field. 
Then the time evolution of the density matrix $\rho$ obeys the master equation, $\frac{d\rho}{dt} = {\cal{L}}\rho$ with the Liouvillian superoperator defined as 
\begin{equation}\label{master equation}%
{\cal{L}}\rho = -i [\hat{H}, {\rho}] + \frac{\kappa}{2} \mathcal
{\cal{D}}[\hat{a}]\rho + \frac{\gamma}{2} \mathcal
{\cal{D}}[\hat{\sigma}_{13}]\rho + \frac{\gamma}{2} \mathcal
{\cal{D}}[\hat{\sigma}_{23}]\rho ,
\end{equation}
where $\mathcal {D}[\hat{o}]\rho=
2\hat{o} {\rho} \hat{o}^\dag - \hat{o}^\dag \hat{o}{\rho} - {\rho}
\hat{o}^\dag \hat{o}$ denotes the Lindblad type of dissipation. Thus, the steady-state intracavity photon number ${n}_s = {\rm Tr}(\hat{a}^\dag \hat{a}\rho_s)$ is evaluated by solving the steady-state master equation with ${\cal L}\rho_s = 0$. It is straightforward to obtain the cavity transmission $T_a= n_s/n_0$ with $n_0=(\eta/\kappa)^2$ being the bare cavity photon number. In addition, the general time interval $\tau$ dependence second-order correlation function for isolated photons is defined as
\begin{align}
g^{(2)}(\tau)=\frac{{\rm Tr}[\hat{a}^\dagger(t)\hat{a}^\dagger(t+\tau)
\hat{a}(t+\tau)\hat{a}(t)\rho_s]}{{\rm Tr}[\hat{a}^\dagger(t)
\hat{a}(t)\rho_s]^2}.
\end{align}
For $\tau=0$, $g^{(2)}(\tau)$ is directly reduced to the equal time  second-order correlation function $g^{(2)}(0)$. As a result, the photon quantum statistics for PB should satisfy both the subPoissonian statistics with $g^{(2)}(0)<1$ and  photon antibunching with $g^{(2)}(0)<g^{(2)}(\tau)$, simultaneously. Without loss of generality, the strong PB is defined as $g^{(2)}(0)<0.01$, where photon antibunching is confirmed by using the quantum regression theorem~\cite{Tang19}.

In our numerical simulation, we take the cavity decay rate $\kappa=2\pi\times160$ kHz, which has been realized in recent experiment of superradiance for $^{87}$Sr clock transition~\cite{Norciae1601231}, the atomic spontaneous decay rate $\gamma=2\pi\times7.5$ kHz for the long-lived excited state of $^3P_1$, the weak cavity driven strength $\eta/\kappa=0.1$, and the single atom-cavity coupling $g/\kappa=4$, for which coupling strength on a forbidden transition has demonstrated capability in current experiments for single atoms trapped in the cavity with the quantization length of the cavity around a few millimeters~\cite{PhysRevLett.118.263601,Birnbaum2005}. Therefore, the free parameters in our system are reduced to cavity-light detuning $\Delta_c$, Rabi frequency of the control field $\Omega$, and optical Stark shift $U_0$  by continuously tuning the sign and strength of Zeeman splitting $\hbar\Delta$ in experiment~\cite{PhysRevLett.118.263601}. Limited by the validity of our model, the numerical results presented below exhibit the parameter space $0\leq U_0/g\leq 4$,  corresponding the tunable magnetic field around a few tens of Gauss. Finally, we point out that our scheme should also be applicable to other alkaline-earth-metal atoms, e.g., Yb atom~\cite{PhysRevLett.120.083601}, which also possess the long-lived excited states.

\begin{figure}[ht]
\includegraphics[width=1\columnwidth]{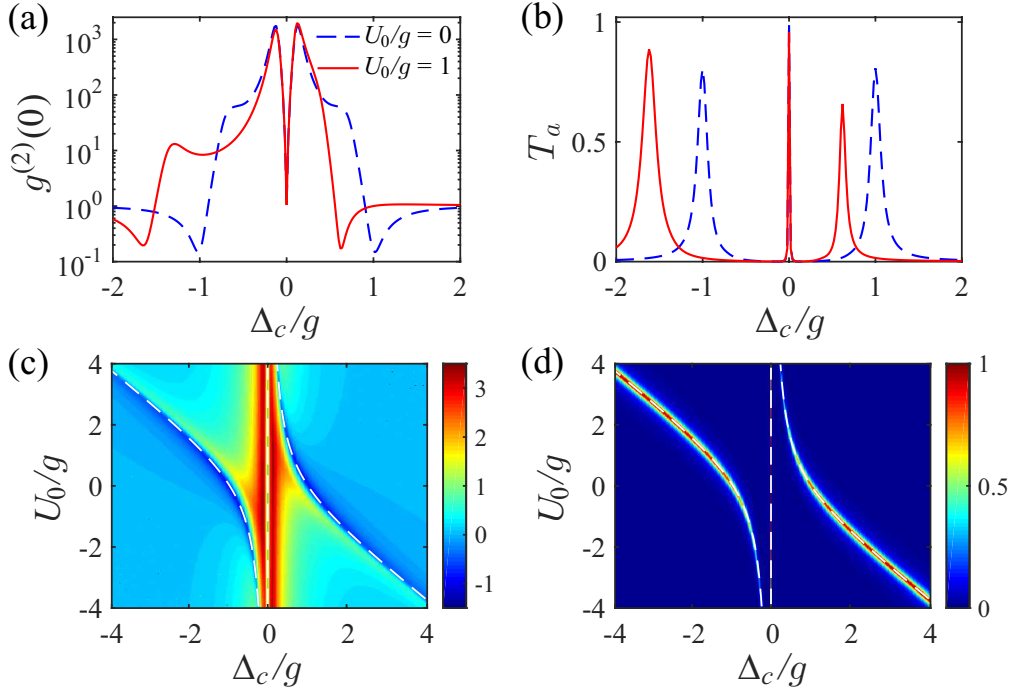}
\caption{(a)  Second-order correlation function $g^{(2)}(0)$ and (b) cavity transmission $T_a$ as a function of cavity-light detuning $\Delta_c$ for $U_0/g$= 0 and 1. (c) $g^{(2)}(0)$ and (d) $T_a$ as a function of $\Delta_c$ and optical Stark shift $U_0$. The white dashed line shows the corresponding dressed state splitting for the three branches of the energy spectrum. The colors with blue-red gradient shading indicate the values of log$[g^{(2)}(0)]$ in (c) and $T_a$ in (d).} \label{gamma}
\end{figure}%

In the weak control field limit, e.g., $\Omega/g =0.01$, we plot the second-order correlation function at zero time delay $g^{(2)}(0)$ and the cavity transmission $T_a$ as a function of cavity-light detuning $\Delta_c$ for different optical Stark shift $U_0$, as shown in Figs.~\ref{gamma}(a) and \ref{gamma}(b). As can be seen, a symmetry structure for $g^{(2)}(0)$ with two minimum values appearing at the red and blue sidebands with $\Delta_c=\delta_{1,\pm}$ is observed in absence of the Stark shift. Besides, a narrow EIT transmission window appears at atomic dark-state resonance with $\Delta_c=\delta_{1,0}$, which emerges due to the quantum interference between the cavity field and the control light. In particular, the nearly $100 \%$ cavity transmission maintains a coherent quantum statistic by keeping $g^{(2)}(0)=1$ when $U_0=0$. As expected, the symmetry structures for $g^{(2)}(0)$ and $T_a$ between the red and blue sidebands are broken when the optical Stark shift $U_0$ are introduced. However, the cavity transmission $T_a$ and photon quantum statistics $g^{(2)}(0)$ are insensitive to $U_0$, although the Stark shift obviously changes the vacuum-Rabi splitting at the red and blue sidebands of the system.

Figures.~\ref{gamma}(c) and \ref{gamma}(d) illustrate the results for log[$g^{(2)}(0)$] and $T_a$ as a function of $U_0$ and $\Delta_c$. It is shown that $g^{(2)}(0)$ and $T_a$ both depend on the signs of $U_0$ and $\Delta_c$, and the smaller value of $T_a$ corresponds to the smaller $g^{(2)}(0)$ at the red (blue) sideband of vacuum-Rabi resonance for positive (negative) $U_0$. The white dashed lines show the corresponding dressed-state splitting of the three branches by analytically calculating the energy spectrum, which are highly consistent with the numerical results by solving the master equation in Eq.(\ref{master equation}) as well. In contrast to the spin-$1/2$ model~\cite{Tang19}, the $g^{(2)}(0)$ does not exhibit a significant reduction with the enhanced PB, even for a large $|U_0|$. It is understood that the quantum interference mechanism for realizing strong PB is absent in our cavity-EIT model at the vacuum-Rabi splitting ($\Delta_c=\delta_{1,\pm}$), which reveals that the Stark shift enhanced energy-spectrum anharmonicity is not sufficient to guarantee the strong PB.

\begin{figure}[ht]
\includegraphics[width=1\columnwidth]{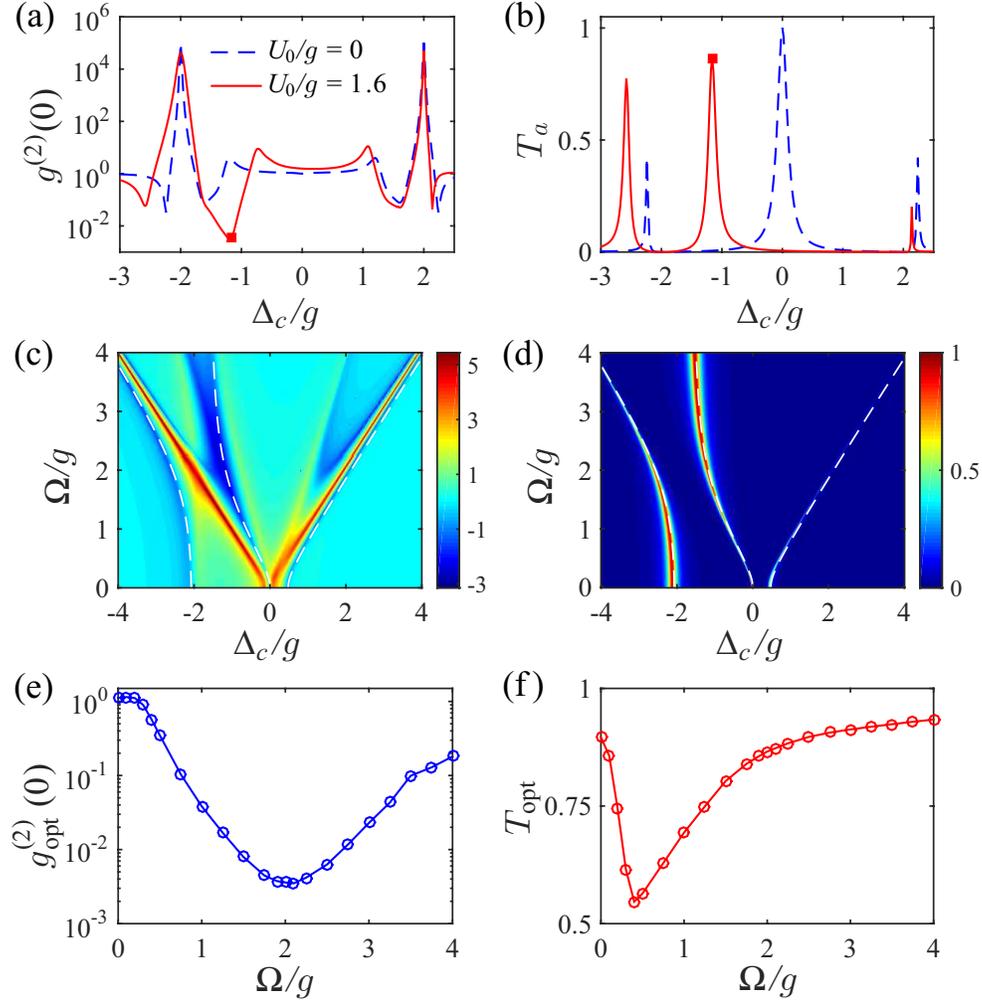}
\caption{ (a) $g^{(2)}(0)$ and (b) $T_a$ as a function of $\Delta_c$ with $\Omega/g=2$ for $U_0/g$=0 and 1.6. The solid blocks mark the minimum value of $g^{(2)}(0)$ for $U_0/g$=1.6 and the corresponding $T_a$. Log$[g^{(2)}(0)]$ (c) and $T_a$ (d) as a function of $\Delta_c$ and $\Omega$. The white dashed line shows the corresponding dressed state splitting for the three branches of the energy spectrum. The optimal $g^{(2)}_{\rm{opt}}(0)$ (e) and the corresponding cavity transmission $T_{\rm{opt}}$ (f) as a function of $\Omega$. In (c)-(d), the other parameter is fixed at $U_0/g=1.6$.} \label{U0}
\end{figure}%

To proceed further, we study the photon quantum statistics $g^{(2)}(0)$ and cavity transmission $T_a$ in a large control field by fixing $\Omega/g=2$. As shown in Figs.~\ref{U0}(a) and \ref{U0}(b), compared with the weak control field case, we find that $T_a$ decreases rapidly with the increasing $\Omega$ at the red and blue sidebands of vacuum-Rabi resonance for $U_0/g=0$ (blue dashed line), although the corresponding $g^{(2)}(0)$ has a slight decrease due to the large control field enhanced cavity-EIT nonlinearity. But, for nonzero Stark shift with $U_0/g=1.6$ (red solid line), we notice that the photon quantum statistics host a strong PB with $g^{(2)}(0)=3.6\times10^{-3}$ at atomic quasi-dark resonance with $\delta_{1,0}= -1.16g$ in Fig.~\ref{U0}(a). In particular, the output of the cavity filed still exhibits a high transmission $T_a$, which is apparently larger than the value of $T_a$ at the red sideband of vacuum-Rabi splitting [Fig.~\ref{U0}(b)]. Therefore, the realization of strong PB by utilizing atomic quasi-dark resonant with a high transmission is of great benefit to generate an ideal single photon source in cavity-EIT. We should note that the predicted narrow transmission transparency window employing advantages of cavity-EIT may facilitate the studies of high quality single atom transistors and all-optical switching in single photon levels~\cite{PhysRevLett.78.3221,Albert2011,Nphoton.6.605,Chen768}.

Figures.~\ref{U0}(c) and~\ref{U0}(d) display the contour plots of $g^{(2)}(0)$ and $T_a$ on the $\Delta_c$-$\Omega$ parameter plane for $U_0/g=1.6$. As can be seen, atomic quasi-dark resonance with strong PB is observed by tuning the Rabi frequency $\Omega$ of the control field. The dressed state splitting for the three branches exhibits the large nonlinear shifts with the increasing $\Omega$, which also agrees with the analytically asymmetric energy splitting for the first manifold in energy spectrum (white dashed lines). In order to quantitatively characterize the quantum statistics for the cavity field, we plot the minimum $g^{(2)}_{\rm{opt}}(0)=\min[g^{(2)}(\Delta_c)]$ and the corresponding cavity transmission $T_{\rm{opt}}$ at the same value of the cavity-light detuning versus $\Omega$, as displayed in figs.~\ref{U0}(e) and~\ref{U0}(f). It is clear that both $g^{(2)}_{\rm{opt}}(0)$ and $T_{\rm{opt}}$ possess a dip feature, corresponding the local minimum with $g^{(2)}(0)=3.5\times10^{-3}$ and $T=0.55$ for different $\Omega$. Interestingly, the optimal PB with the optical control field $\Omega/g=2.1$ corresponds to a large cavity transition up to $T=0.87$, which is crucial for applications of high-quality single photon sources.

\begin{figure}[ht]
\includegraphics[width=1\columnwidth]{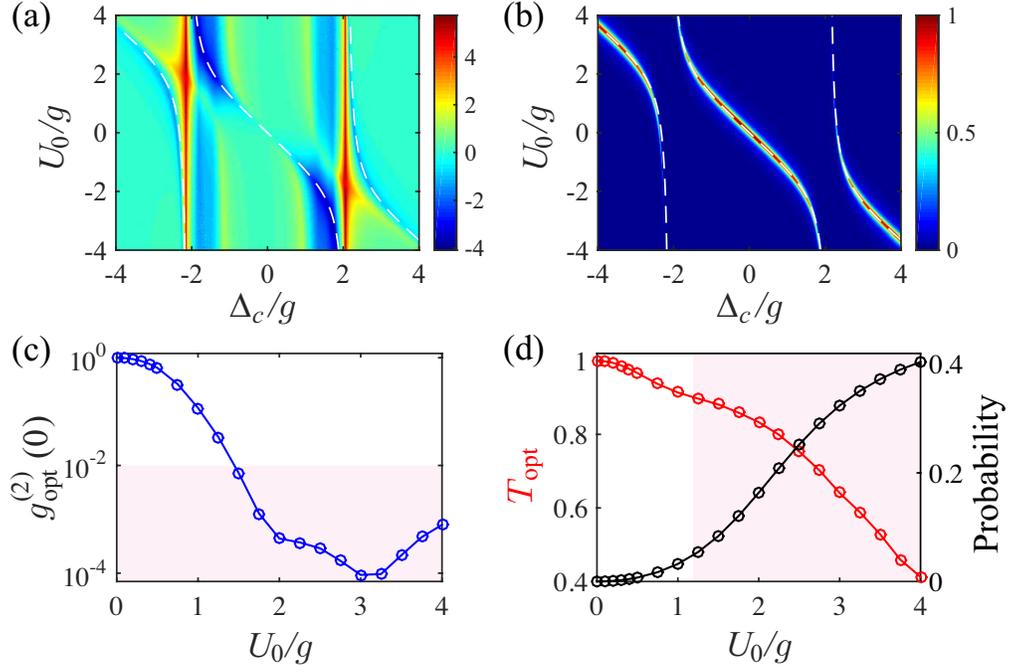}
\caption{Contour plots of log$[g^{(2)}(0)]$ (a) and cavity transmission (b) as a function of $\Delta_c$ and $U_0$. The white dashed line shows the dressed state splitting for the three branches of energy spectrum. The optimal $g^{(2)}_{\rm{opt}}(0)$ (c) and the corresponding $T_{\rm{opt}}$ [the red curve] (d) as a function of $U_0$, and the light pink area exhibit the strong PB regime and the corresponding $T_a$, respectively. The black curve in (d) shows the corresponding occupy probability of atomic excited state $|3\rangle$. Here, the  classical control field is fixing at the optimal value with $\Omega/g=2.1$.} \label{omega}
\end{figure}%

To gain more insight into the PB in the cavity-EIT, we further calculate the  $g^{(2)}(0)$ and $T_a$ in the $\Delta_c$-$U_0$ parameter plane by fixing the optimal control field $\Omega/g=2.1$, as shown in Figs.~\ref{omega}(a) and~\ref{omega}(b). Compared with the results of the weak control field shown in Figs.~\ref{gamma}(c) and \ref{gamma}(d), the photon quantum statistics and cavity transition are highly sensitive to the strength and sign of the optical Stark shift for the large control field. In particular, the strong PB occurs at the middle branch accompanied with atomic quasi-dark resonance. With the increasing of $U_0$, the optimal $g^{(2)}_{\rm{opt}}(0)$ displays a dip structure, which is dramatically decreasing from one to a local minimum with a optimal $g^{(2)}_{\rm{opt}}(0)=9.0\times10^{-5}$ at $U_0/g=3.0$ [Fig.~\ref{omega}(c)]. The typical feature of $T_{\rm{opt}}$ is gradually decreasing as a function of $U_0$, corresponding the probability of the excited state $|\beta|^2$ in Eq.~(\ref{DSP}) that is gradually growing at the atomic quasi-dark resonance [Fig.~\ref{omega}(d)].  We also checked that the photon antibunching $g^{(2)}(0)$ gradually increases with the driven amplitude $\eta$ of the input laser, albeit the steady-state intracavity photon number $n_s$ is obviously growing.

Remarkably, the optimal $g^{(2)}(0)$ for strong PB emerging at atomic quasi-dark resonance is more than four orders of magnitude smaller than the case of coherent EIT transition in the absence of $U_0$. We emphasize that the strong PB is contributed by combination of the Stark shift enhanced energy-spectrum anharmonicity and EIT hosted large nonlinearity at atomic quasi-dark state resonance. In addition, the strong PB with a large cavity transmission exists in a large parameter region, even deviating from the optimal parameters [the light pink area in Fig.~\ref{omega}(c)], which could facilitate the experimental feasibility for studying the exotic nonclassical quantum states in cavity-EIT.

Compared with the Stark shift term of $U_0 \hat{a}^\dag\hat{a} \hat{\sigma}_{11}$ in Eq.~(\ref{Hamiltonian}), the reduced Stark shift for the typical scheme of a single four-level atom-cavity-EIT~\cite{PhysRevA.61.011801,Rebic_1999,rebic2002photon,bajcsy2013photon} takes the form of $U_0 \hat{a}^\dag\hat{a} \hat{\sigma}_{22}$, in which the cavity field drives the far-resonance atomic $|2\rangle \leftrightarrow |4\rangle$ transition. Therefore, the realization of PB in these proposals~\cite{PhysRevA.61.011801,  Rebic_1999, bajcsy2013photon,rebic2002photon, PhysRevA.65.043806, Greentree_2000} can be ascribed to the large nonlinearity from generating the Kerr effect. However, the strong PB in our model with existing the optimal optical Stark shift and the classical control field announces that the mechanism for generating strong PB cannot be completely ascribe to the Kerr nonlinearity in cavity-EIT~\cite{schmidt1996giant,Imamoglu97}.

\begin{figure}[ht]
\includegraphics[width=1\columnwidth]{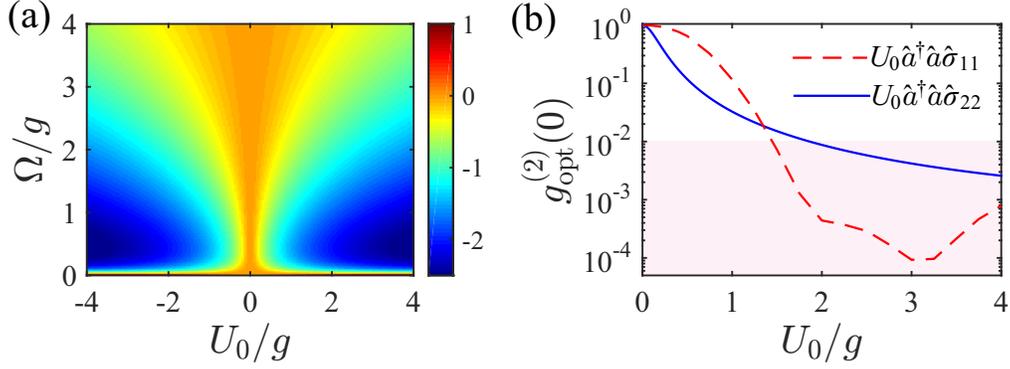}
\caption{(a) Contour plots of log$[g^{(2)}(0)]$ for the earlier proposals in Ref.~\cite{PhysRevA.61.011801,Rebic_1999,rebic2002photon,bajcsy2013photon} on the $U_0$-$\Omega$ parameter plane with $\Delta_c/g=0$.  The
light pink area exhibiting the strong PB regime with ($g^{(2)}(0)<0.01$) is a guide for the eye. (b) The optimal $g^{(2)}_{\rm{opt}}(0)$ in the presence of Stark shift $U_0 \hat{a}^\dag\hat{a} \hat{\sigma}_{11}$ (dashed  line) with $\Omega/g=2.1$ and $U_0 \hat{a}^\dag\hat{a} \hat{\sigma}_{22}$ (solid line) with $\Omega/g=0.44$  as a function of $U_0$, respectively.}\label{Kerr}
\end{figure}%

To further quantitatively distinguish the essential differences of the two proposals, we calculate $g^{(2)}(0)$ in the presence of Stark shift term of $U_0 \hat{a}^\dag\hat{a} \hat{\sigma}_{22}$ in the $U_0$-$\Omega$ parameter plane, as displayed in Fig.~\ref{Kerr}(a). Here, the cavity-light detuning is fixed at $\Delta/g=0$, since the optimal PB occurred at the position of atomic dark-state resonance for atom-cavity EIT. For a  fixed $|U_0|$, the minimum photon antibunching amplitude of $g^{(2)}(0)$ corresponds to the optimal control field with $\Omega/g=0.44$. It is clear that  the optimal $g^{(2)}_{\rm{opt}}(0)$ is monotonically decreased with the increasing of the effective Kerr nonlinearity by tuning $U_0$. In order to compare the PB between our proposal and earlier schemes~\cite{PhysRevA.61.011801,Rebic_1999,rebic2002photon,bajcsy2013photon} more intuitively, the optimal $g^{(2)}_{\rm{opt}}(0)$ as a function of $U_0$ is illustrated, as shown in Fig.~\ref{Kerr}(b). As can be seen, the optimal value of $g^{(2)}(0)$ in our scheme is much less than the previous proposals from hosting a wide range parameter regime of strong PB. In particular, $g^{(2)}(0)$ in our scheme can be reduced more than $45$ times in magnitude at a moderate Stark shift with $U_0/g=3.0$,  where the strong antibunching photons are ascribed to the significantly enhanced energy-spectrum anharmonicity [Fig.~\ref{scheme}(c)] induced by $U_0$.

Furthermore, the essential difference between the two schemes for generating PB can be readily understood by comparing the energy spectrum of the system. For the single four-level atom-cavity-EIT in Refs.~\cite{PhysRevA.61.011801,Rebic_1999,rebic2002photon,bajcsy2013photon}, we find that the first dressed state in the energy spectrum is immune to the changing of $U_0$. Therefore, the vacuum-Rabi splitting  $\delta_{1,\pm}= \pm \sqrt{g^2+\Omega^2}$ and $\delta_{1,0}=0$ for the three branches are all independent of $U_0$. From this perspective, the Stark shift enhanced vacuum-Rabi splitting in our proposal is equivalent to enlarging the effective single atom-cavity coupling.  From the experimental point of view, the realization of strong PB beyond the strong optical Stark shift in our scheme could obviously improve the accessibility and controllability in the experiment. Therefore, our proposal can be used to realize a high-quality single photon source for a single atom-cavity-EIT mediated by a moderate optical Stark shift. We should note that the advantage of generating PB based on Kerr nonlinearity is  using an ultracold ensemble coupled to the cavity~\cite{schmidt1996giant,Imamoglu97}, where the giant Kerr nonlinearity could be emerged on the atomic dark state resonance of cavity-EIT.
 
\section{Conclusions}
In summary, we propose an experimental scheme for generating PB in optical Stark shift mediated-cavity-EIT by employing the advantages of energy-level structures in single alkaline-earth-metal atoms. By utilizing atomic quasi-dark state resonance, strong PB with second-order correlation function and high cavity transmission are achieved by tuning the optimal Stark shift and control field. The mechanism in our model is owe to the combination of EIT hosted large nonlinearity and Stark shift enhanced energy-spectrum anharmonicity, which provides new insights to the PB mechanism in cavity-EIT beyond the giant Kerr nonlinearity. Importantly, all parameters for our numerical simulation are based on the current experimental capabilities. Compared with the traditional single four-level atom-cavity-EIT systems~\cite{PhysRevA.61.011801,Rebic_1999,rebic2002photon,bajcsy2013photon}, the strong PB with $g^{(2)}(0)=9\times 10^{-5}$ can be achieved at a moderate Stark shift, which could significantly facilitate the realization of strong PB beyond the strong atom-cavity coupling. 

Meanwhile, our scheme has the following advantages in contrast to the Stark shift mediated Jaynes-Cummings model~\cite{Tang19}. 
(i) Our scheme does not require the quantum interference conditions, which rely on the precise control of the phase and the coupling strength of an external microwave field simultaneously. 
Indeed, the quantum interference conditions are difficult to implement for a complex multilevel atom-cavity system. 
(ii) By utilizing the atomic quasi-dark-state resonance of cavity-EIT, the large cavity transmission can be expected, which is insensitive to the inevitable spontaneous emission of excited states. 
(iii) In our configuration, the photon antibunching amplitude of $g^{(2)}(0)$ can be reduced more than one order of magnitude compared with the proposal of quantum interference at a moderate Stark shift. 
In fact, we should emphasize that the multiphoton ($n\geq3$) excitations are hard to be suppressed for the quantum interference mechanism, although the two-photon excitation is completely eliminated. 
Finally, our experimental proposal for generation of high-quality single photon sources could implicate exciting opportunities for potential applications in quantum information science and quantum metrology~\cite{RevModPhys.90.045005}.

\section*{Fundings}

This work is supported by the National Natural Science Foundation of China (11804409,  11874433, 12025509, 11874434), the National Key R$\&$D Program of China (2018YFA0307500), the Fundamental Research Funds for the Central Universities (Grant No. 18lgpy80), and the Key-Area Research and Development Program of Guang-Dong Province under Grants No. 2019B030330001.

\section*{Disclosures}

The authors declare no conflicts of interest.

%
\end{document}